# Performance Variance of Low Noise Resonant Capacitance Bridges While Replacing their Ungapped MnZn Ferrite Cores

S. Saraf[1], S. Buchman[2], C.Y. Lui[1], S. Wang[1,3], J. Lipa[2]

[1]*SN&N Electronics, Inc., 1846 Stone Avenue, San Jose, CA 95125 USA*
[2]*Hansen Experimental Physics Laboratory, Stanford, CA 95038 USA*
[3]*Hainan Tropical Ocean University, Sanya, 572022 China*

The author to whom correspondence should be addressed: **saraf@snnelectronics.com**

## ABSTRACT

Precision AC resonant capacitance bridges, with a planar printed circuit board transformer using an ungapped MnZn ferrite core, have shown excellent noise performance in high-precision measurements. As many applications use an ensemble of bridges, consistency of performance is critical to the functionality of the systems. If part of the same manufacturing batch and scaled to the same frequency, the noise performance of the ferrite cores at 293 K has a weighted mean variance of < 0.3%, with all the cores within a ± 1% band. At 140 K the weighted mean variance is < 0.6% and the 90% inclusion band is ± 1.5%. Ten cores of SIFERRIT Material N41, from TDK Ferrites Accessories, were tested at room temperature, 293 K, and in a liquid nitrogen dewar at 140 K. Fitted to a parabolic function, to + 3dB on both frequency sides, the weighted mean of the noise minima at 293 K was 0.3027 aF/$\sqrt{\text{Hz}}$, with a variance of 0.0012 aF/$\sqrt{\text{Hz}}$ - spanning a frequency range of 85.0 ± 2.7 kHz. Scaled to 100 kHz, the weighted mean and variance were 0.2788 aF/$\sqrt{\text{Hz}}$ and 0.0008 aF/$\sqrt{\text{Hz}}$ . Corresponding noise values at 140 K were 0.1815 aF/$\sqrt{\text{Hz}}$ with a variance of 0.0016 aF/$\sqrt{\text{Hz}}$ for the range of 152.7 ± 6.3 kHz, and 0.1835 aF/$\sqrt{\text{Hz}}$ with a variance of 0.0011 aF/$\sqrt{\text{Hz}}$ when scaled to 150 kHz. Scaled for resonant frequency and temperature these results are consistent with a previous measurement of another ungapped core (same supplier and type, different manufacturing batch) to within 9% and 6% at 293 K and 140 K respectively.

## I. INTRODUCTION

Several accurate proximity sensing instruments rely on resonant capacitance bridges (RCBs) for measurements at the sub-attofarad precision levels. Here we consider the applications that use multiple RCBs, paired as differential capacitance bridges, that require good reproducibility among sensors - with stable transformers used to enhance the signal-to-noise ratio, thus determining relative sub-nm displacements. See for example references 1, 2, and 3 for the description of such RCBs, with the detailed specifications of the circuits and the test instrumentation used reported in reference 1. These sensors have applications in geodesy[4], tests of the equivalence principle[5], and future space-based gravitational-wave antennas; LISA[6,7], Taiji[8], and TianQin[9].

One possible source of variability between bridges are the MnZn ferrite cores of the transformers, as the manufacturer's specifications are a somewhat lax ± 25% for their inductance value[10]. In this study, we compare the noise performance of 10 RCBs (designated A to J) where the only variables are the transformer SIFERRIT material N41 cores, purchased as a batch from TDK[11]. These cores were installed sequentially in the transformer board stack of the same RCB circuit board and the noise measurements were performed with the same instrumentation set-up. Figure 1 shows the installation of a transformer core onto the RCB board.

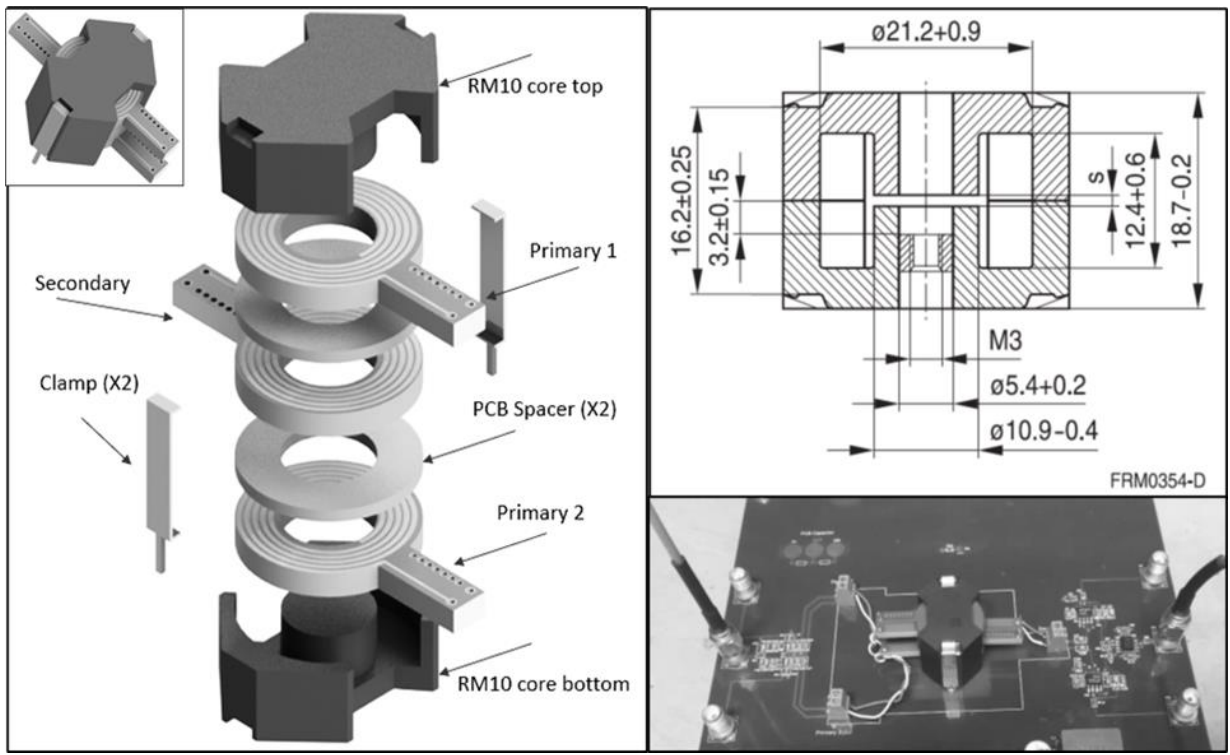

**Figure 1. Photograph of the installation of a core in the RCB circuit.**

For the low-temperature tests each A to J core was cooled in the same liquid nitrogen cryostat, and the same one-hour cool down period was allowed for. Figure 2 shows the schematics of the RCB circuit, with the location of the cores highlighted.

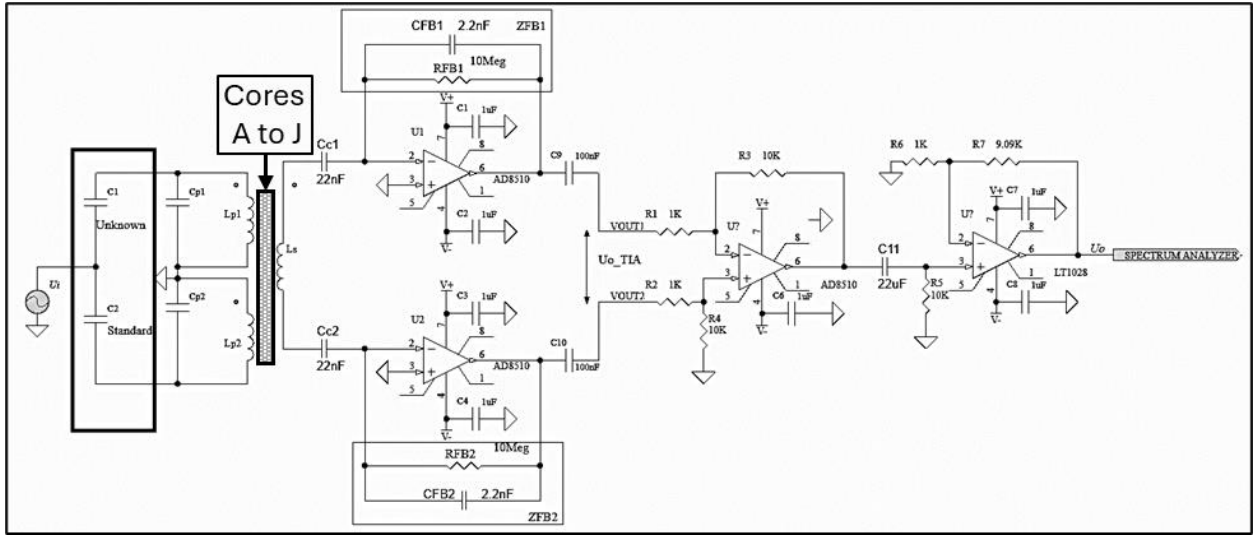

**Figure 2. RCB schematic highlighting the transformer cores.**

Replacing the transformer cores modifies only the inductance $L$ of the circuit, with all other parameters – and in particular the equivalent capacitance $C_{eq}$ - remaining constant. Consequently, variations in the resonant frequency $f_0$ are caused by variations in $L$. The resonant frequency $f_0$ is given by:

$$f_0 = \frac{1}{2\pi\sqrt{C_{eq}L}} \quad \Rightarrow \quad L = \frac{1}{(2\pi f_0)^2 C_{eq}} \tag{1}$$

$S^{1/2}$, the circuit noise in units of $\text{F}/\sqrt{\text{Hz}}$ is given by:

$$S^{1/2} = \frac{1}{U_i} \times \sqrt{\frac{8k_B T}{(2\pi f_0)^3 L(T) Q_{RCB}}} \left(\frac{\text{F}}{\sqrt{\text{Hz}}}\right) \Rightarrow S_C^{1/2} = \frac{1}{U_i} \times \sqrt{\frac{8k_B T C_{eq}}{(2\pi f_0) Q_{RCB}}} \left(\frac{\text{F}}{\sqrt{\text{Hz}}}\right) \tag{2}$$

where $U_i = 1\text{V}$ is the bridge excitation voltage, $Q_{RCB}$ is the quality factor of the RCB, $k_B$ is the Boltzmann constant, and $T = 293\text{K}$ – or $T = 140\text{K}$ for the liquid nitrogen cooled tests - the temperature of the RCB transformer. Data shows that $Q_{RCB}$ is approximately constant for a given temperature (see Table 1 and 2), resulting in the estimate that the noise figure of the RCB is primarily dependent on the inductance $L$ and the resonant frequency $f$ (equation 2). As $L \propto f^{-2}$, $S^{1/2}$ is therefore proportional to the inverse square root of its resonant frequency at constant temperature:

$$S^{1/2} \propto f_0^{-1/2} \tag{3}$$

## II. ROOM TEMPERATURE RESULTS - 293 K

Noise floor measurements were made using a Rohde & Schwarz (R&S) model FSV40-N[12] spectrum analyzer, with a resolution bandwidth $RBW = 100\text{Hz}$, while the bridge input is grounded by $R_{50} = 50\Omega$. The conversion of the RCB noise power $P(\text{dBm})$ from the dBm units measured by the spectrum analyzer to the more intuitive noise units $S_C^{1/2}(\text{aF}/\sqrt{\text{Hz}})$ – to be used throughout – is given by:

$$S_C^{1/2}\left(\text{aF}/\sqrt{\text{Hz}}\right) = \frac{\sqrt{\frac{10^{P(\text{dBm})/10}}{1000}R_{50}(\Omega)}f_{dem}}{G_{diff}G_{ext}\sqrt{RBW(\text{Hz})}f_{conv}(\text{V/aF})} = 366 \times 10^{P(\text{dBm})/20} \quad (4)$$

where $f_{dem} = \sqrt{2}$ accounts for the noise introduced by the demodulation stages in a complete system implementation, $G_{ext} = 10$ is the gain of an external amplification stage that uses a quiet LT1028 operational amplifier, $G_{diff} = 10$ is the gain of an internal capacitively coupled differential amplifier stage, and $f_{conv} = 864\ \mu\text{V/fF}$ (at 1 V excitation) is the measured readout sensitivity conversion factor; see reference 1 for a detailed description. The complete noise spectra in aF/$\sqrt{\text{Hz}}$, with an average of 100 counts, for the ten cores at 293 K is shown in the left of Figure *3* with an expanded view around the noise minima on the right. Note that the overlap in values causes the traces to be hard to separate visually.

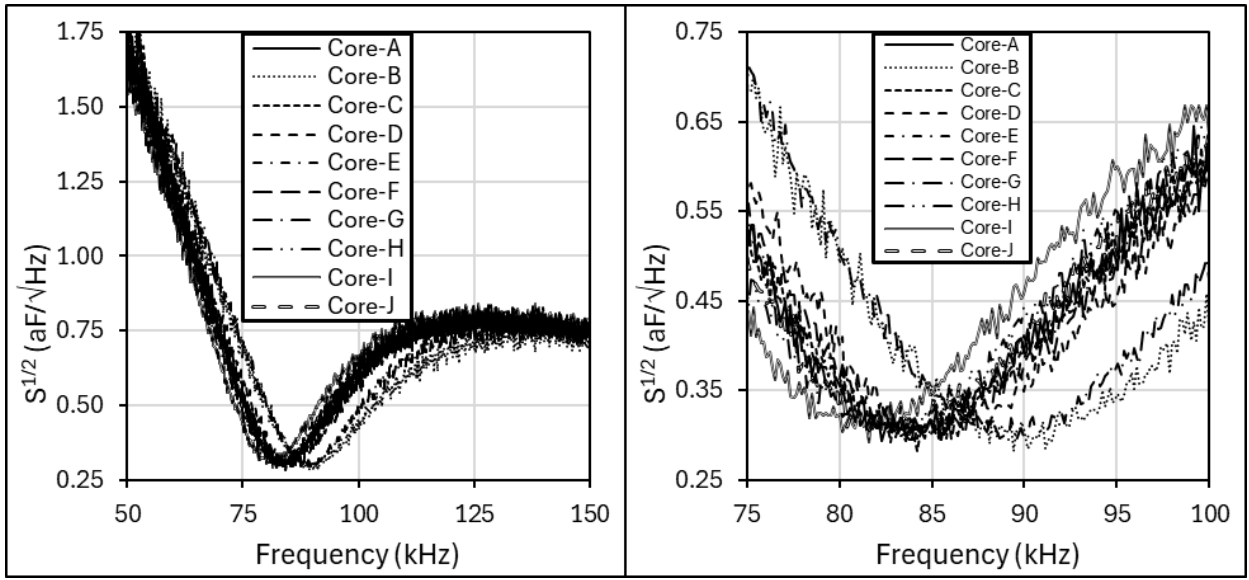

**Figure 3. Left: Noise level $S^{1/2}$ at 293 K for the ten A-J RCB cores. Right: Expanded view around noise minima.**

The minima of the test data and their standard deviation are given in the 'Raw Data' columns of Table 1. For more representative results, individual minimum noise values for each core were calculated by fitting the data to second order polynomials around their lowest points to their + 3 dB ($\times\sqrt{2}$ aF/$\sqrt{\text{Hz}}$) levels for both lower and higher frequencies. As an example, Figure 4 shows the data and parabolic fits for cores A and B, the resultant $S_{min}^{1/2}$ minima of these fits, and their noise values $\Delta S^{1/2}$ calculated from the variance of the parameters of the parabolic fits. The results for the minimum noise of the A-J cores are shown in **Error! Reference source not found.** (with the errors calculated from the parabolic fitting as detailed above) and its overall frequency power trend $S_{min}^{1/2}(f) \propto f^{-.32\pm.03}$. This frequency dependence follows approximately the $\propto f^{-1/2}$ calculation in equation 3, (valid only for identical cores, or same core, at different frequencies) thus reflecting the consistency of parameters of cores from the same batch.

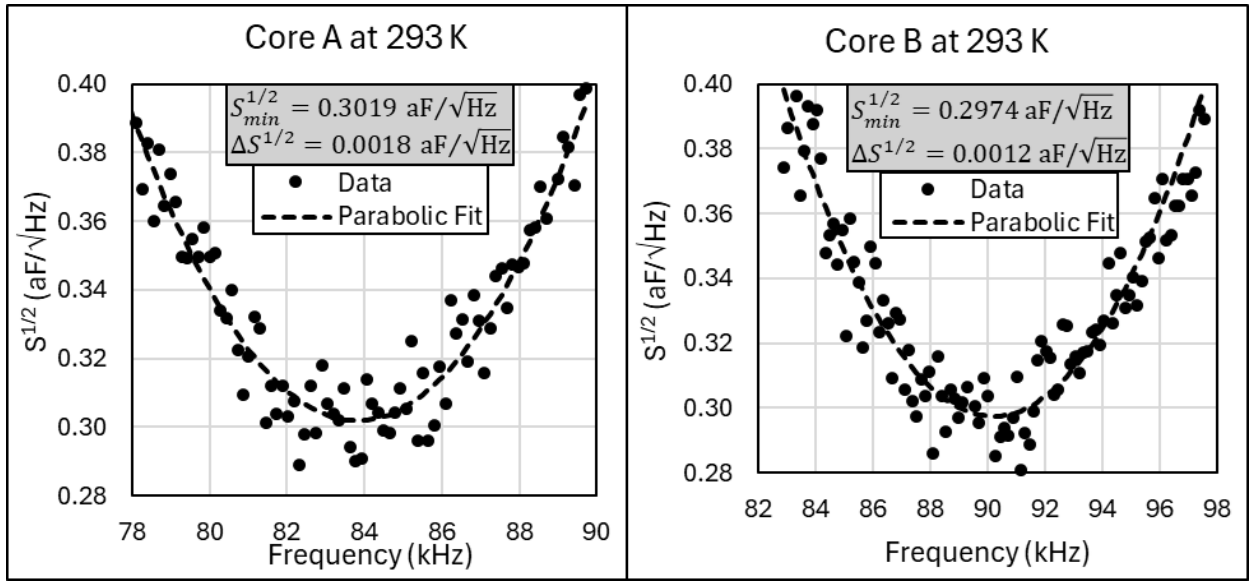

**Figure 4. Parabolic fits to the RCB noise spectra at 293 K for cores A and B.**

The 'Parabolic Fit' columns in Table 1 give the minima $f_{min}$ and $S_{min}^{1/2}$ for the A-J cores, the variance $\Delta S_{min}^{1/2}$ of $S_{min}^{1/2}$, and the quality factor $Q$. The variance $\Delta f_{min}$ of $f_{min}$ is not shown as in all cases $\Delta f_{min} \leq 12$ Hz. $R^2$, the coefficient of determination, is $R^2 > 0.90$ for all cores except core B for which $R^2 = 0.87$. The weighted mean for the ensemble of the A-J cores, using the errors shown in Table 1, is given by $\langle S_{min}^{1/2} \rangle = \langle \Delta S_{min}^{1/2} \rangle^2 \times \sum_{i=A}^{J} \left\{ \left[ 1/\left(S_{min}^{1/2}\right)_i \right] \middle/ \left[ 1/\left(\Delta S_{min}^{1/2}\right)_i^2 \right] \right\}$, with the variance of the weighted mean defined as $\langle \Delta S_{min}^{1/2} \rangle^2 \equiv \left\{ \sum_{i=A}^{J} \left[ 1/\left(\Delta S_{min}^{1/2}\right)_i^2 \right] \right\}^{-1}$.

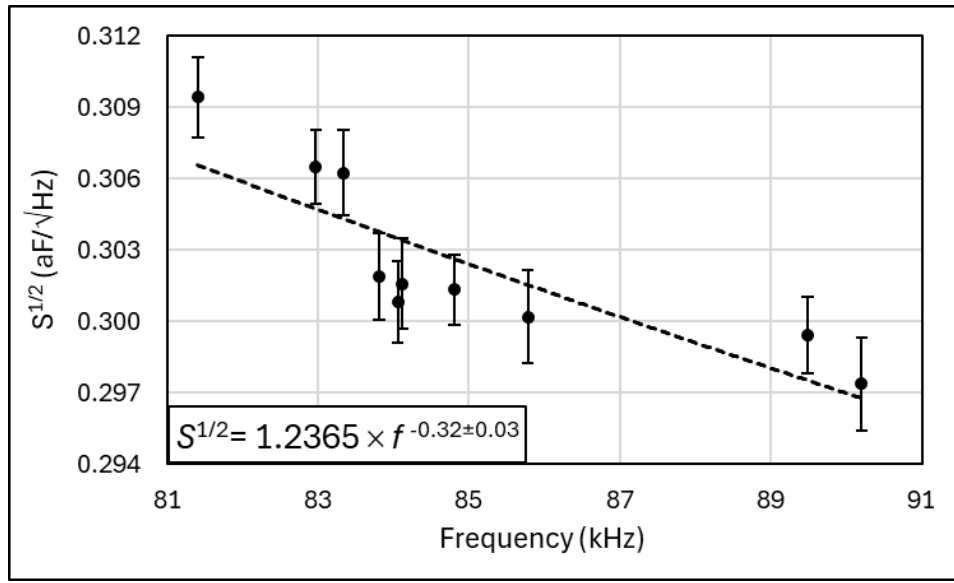

**Figure 5. Noise minima for A-J cores at 293 K derived from the parabolic fits to their spectra and their errors calculated from the variance of the parabolic fits. The dotted line shows the frequency dependence of the noise minima.**

To account for over or under dispersion we calculate the corrected variance in the weighted mean, $\langle \Delta S_{cor}^{1/2} \rangle^2$, given by $\langle \Delta S_{cor}^{1/2} \rangle^2 = \langle \Delta S_{min}^{1/2} \rangle^2 \times \frac{1}{n-1} \sum_{i=A}^{J} \left\{ \left[ \left( S_{min}^{1/2} \right)_i - \langle S_{min}^{1/2} \rangle \right]^2 \big/ \left( \Delta S_{min}^{1/2} \right)_i^2 \right\}$ (see for example reference 13) to obtain:

$$\langle S_{min}^{1/2} \rangle = 0.3027 \, \text{aF}/\sqrt{\text{Hz}} \qquad \langle \Delta S_{cor}^{1/2} \rangle = 0.0012 \, \text{aF}/\sqrt{\text{Hz}} \tag{5}$$

We address the frequency dependence of the noise by scaling $S_{min}^{1/2}$ to 100 kHz, by using equation 3, $S_{100\text{kHz}}^{1/2} = S_{\min}^{1/2} \left[ \frac{f_{res}(\text{kHz})}{100} \right]^{1/2}$, for each resonant frequency $f_{res}$. Using the same statistical approach as above we get for the weighted mean and the corrected variance of $S_{100\text{kHz}}^{1/2}$:

$$\langle S_{100\text{kHz}}^{1/2} \rangle = 0.2788 \, \text{aF}/\sqrt{\text{Hz}} \qquad \langle \Delta S_{100\text{kHz}}^{1/2} \rangle_{cor} = 0.0008 \, \text{aF}/\sqrt{\text{Hz}} \tag{6}$$

Figure 6 and the last column of Table 1 shows the minimum noise data scaled to 100 kHz with the weighted mean (heavy dashed line), its corrected variance (dark shaded area), and the 1% limits (light shaded area between dotted lines) that include all data in within their error bars. In rows 'Weighted mean' and 'Variance of weighted mean' Table 1 gives the results at 293 K for the $S_{\min}^{1/2}$ weighted mean and its corrected variance. For the $f_{min}$, $Q$, and $S_{\min}^{1/2}$ scaled to 100 kHz columns the mean and weighted mean are the averages and standard deviations of the A-J values. The last row of Table 1 gives the result from reference 1 scaled to 100 kHz, that used the same RCB circuit and a similar ungapped core, from the same manufacturer, though from a different batch.

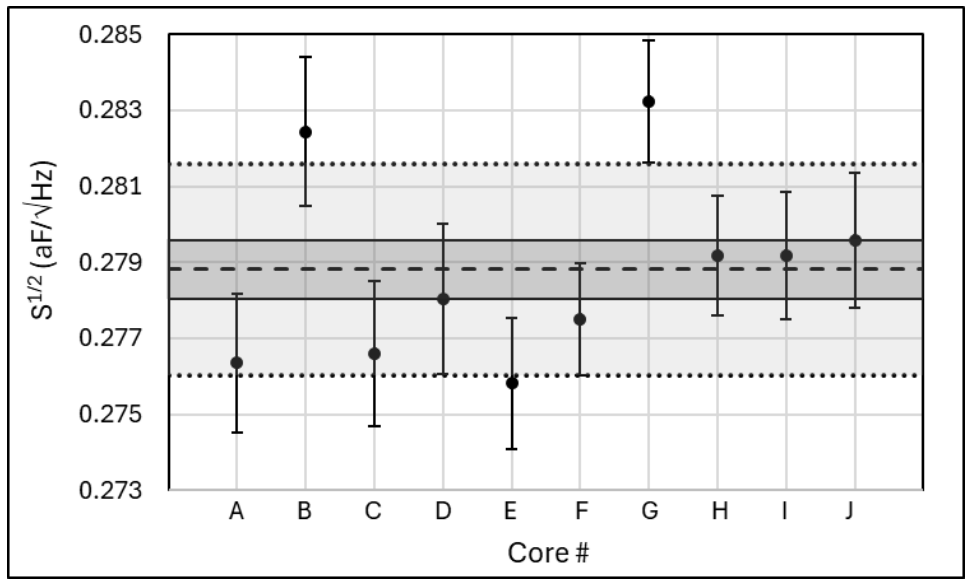

**Figure 6. 293 K data scaled to 100 kHz with the weighted mean shown by dashed line, variance of weighted mean dark shaded, and 1% limits light shaded.**

**Table 1. Summary of 293 K data for the A-J cores.**

| Core # | Raw Data | | Parabolic Fit | | | | Scaled to 100kHz |
|---|---|---|---|---|---|---|---|
| | $S_{min}^{1/2}$ (aF/√Hz) | $\Delta S_{min}^{1/2}$ (aF/√Hz) | $f_{min}$ (kHz) | $S_{min}^{1/2}$ (aF/√Hz) | $\Delta S^{1/2}$ (aF/√Hz) | $Q$ | $S_{min}^{1/2}$ (aF/√Hz) |
| A | 0.289 | 0.009 | 83.81 | 0.3019 | 0.0018 | 6.3 | 0.2764 |
| B | 0.278 | 0.010 | 90.20 | 0.2974 | 0.0020 | 6.5 | 0.2824 |
| C | 0.288 | 0.009 | 84.12 | 0.3016 | 0.0019 | 6.3 | 0.2766 |
| D | 0.289 | 0.009 | 85.78 | 0.3002 | 0.0020 | 6.5 | 0.2780 |
| E | 0.290 | 0.010 | 84.06 | 0.3008 | 0.0017 | 6.3 | 0.2758 |
| F | 0.292 | 0.007 | 84.81 | 0.3013 | 0.0015 | 6.4 | 0.2775 |
| G | 0.289 | 0.007 | 89.49 | 0.2994 | 0.0016 | 6.2 | 0.2832 |
| H | 0.299 | 0.008 | 82.97 | 0.3065 | 0.0016 | 6.3 | 0.2792 |
| I | 0.299 | 0.009 | 81.41 | 0.3094 | 0.0017 | 6.5 | 0.2792 |
| J | 0.297 | 0.009 | 83.34 | 0.3063 | 0.0018 | 6.4 | 0.2796 |
| **Weighted mean** | | | **85.00** | **0.3027** | | **6.37** | **0.2788** |
| **Variance of weighted mean** | | | **2.66** | **0.0012** | | **0.11** | **0.0008** |
| *Ref. 1 scaled to 100 kHz* | | | | | | 8.3 | 0.2548 |

## III. LOW TEMPERATURE RESULTS - 140 K

For the low temperature tests, we used the same cores A-J and the same RCB board and test instrumentation as for the room temperature measurements described above in section II. Each core was mounted on the RCB board, and the assembly was cooled for one hour in liquid nitrogen vapor reaching an equilibrium temperature of 140 K. The data analysis and display follow the same procedure as in section II, starting with $S^{1/2}$, the noise spectra in aF/√Hz, with an average 100 counts for the ten RCB cores A-J, is shown in the left panel of Figure 7 with an expanded view around the noise minima displayed in the right panel.

To determine their minima, we fit the 140 K noise spectra to second order polynomials around their lowest power readings to their + 3 dB level (×√2 aF/√Hz). Figure 8 shows the fits for cores A and B and the resultant $S_{min}^{1/2}$ minima and $\Delta S^{1/2}$ noise values. For all 10 cores $\Delta f_{min}$, the variance of the fitted frequency, is $\Delta f_{min} \leq 18$ Hz, while $R^2$, the coefficient of determination, is $R^2 > 0.92$.

As for the room temperature data, the results for the minimum noise of the A-J cores at 140 K are shown in Figure 9. A power fit to the $S_{min}^{1/2}$ data in Figure 9, dotted line, shows the dependence on the resonant frequency $S_{min}^{1/2}(f) \propto f^{-.52\pm.03}$. The weighted mean and its corrected variance for the ensemble of A-J cores at 140 K are given by:

$$\langle S_{min}^{1/2} \rangle = 0.1815 \text{ aF}/\sqrt{\text{Hz}} \qquad \langle \Delta S_{cor}^{1/2} \rangle = 0.0016 \text{ aF}/\sqrt{\text{Hz}} \qquad (7)$$

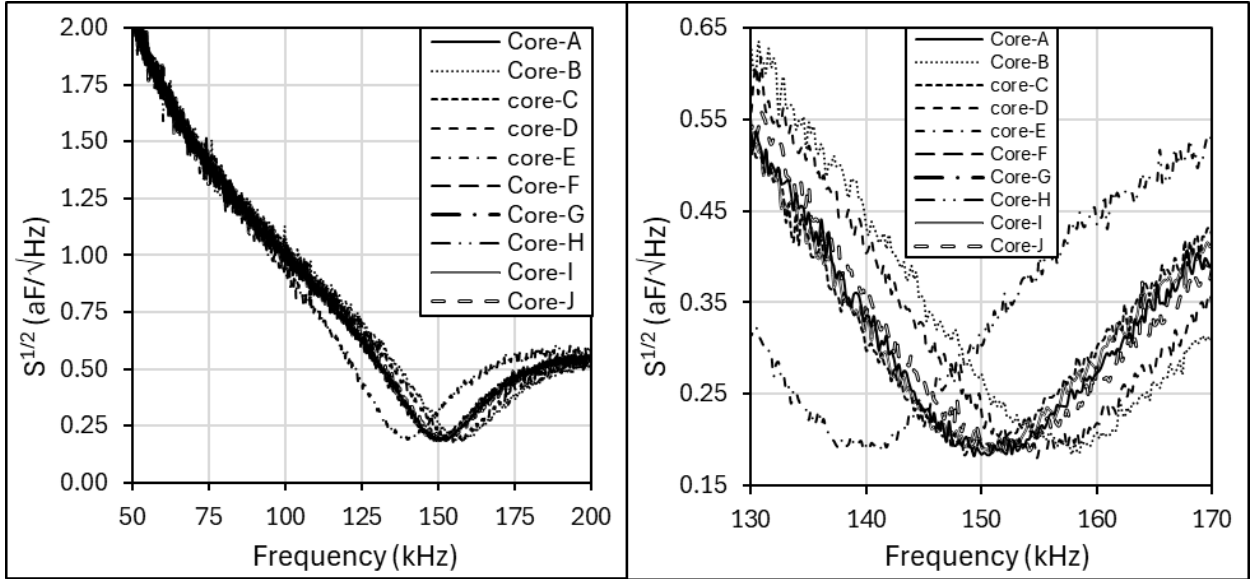

**Figure 7. Left: Noise level $S^{1/2}$ at 140 K for the ten A-J RCB cores. Right: Expanded view around noise minima.**

Using equation 3 we scale for the frequency dependence of the noise at 140 K by scaling $S^{1/2}_{fres}$ to a nominal 150 kHz for each of the A-J cores and obtain:

$$\langle S^{1/2}_{150\text{kHz}} \rangle = 0.1835 \text{ aF}/\sqrt{\text{Hz}} \qquad \langle \Delta S^{1/2}_{150\text{kHz}} \rangle_{cor} = 0.0011 \text{ aF}/\sqrt{\text{Hz}} \qquad (8)$$

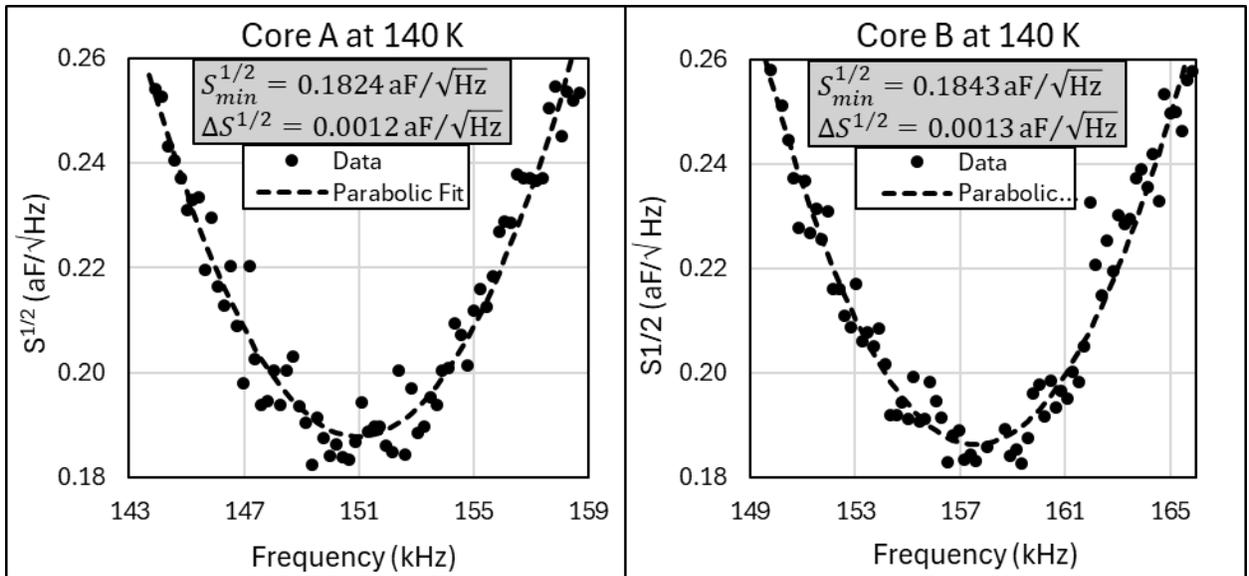

**Figure 8. Parabolic fits to the RCB noise spectra at 140 K for cores A and B.**

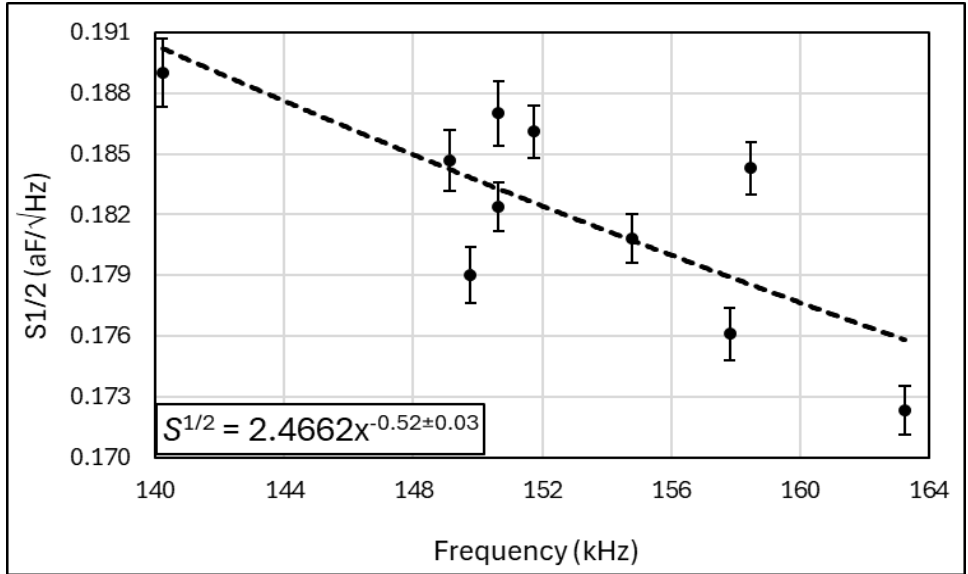

**Figure 9.** Noise minima for A-J cores at 140 K derived from the parabolic fits to their spectra and their errors calculated from the variance of the parabolic fits. The dotted line is the frequency dependence of the minima.

Figure 10 shows the 140 K minimum noise data scaled to 150 kHz (with markings corresponding to those in Figure 6), while Table 2 gives a summary of the results at 140 K with weighted means and variances of weighted means similarly to the notations in Table 1 and includes the results from reference 1 by the same authors using the same circuit and an ungapped core from the same manufacturer, though from a different batch and at 120 K.

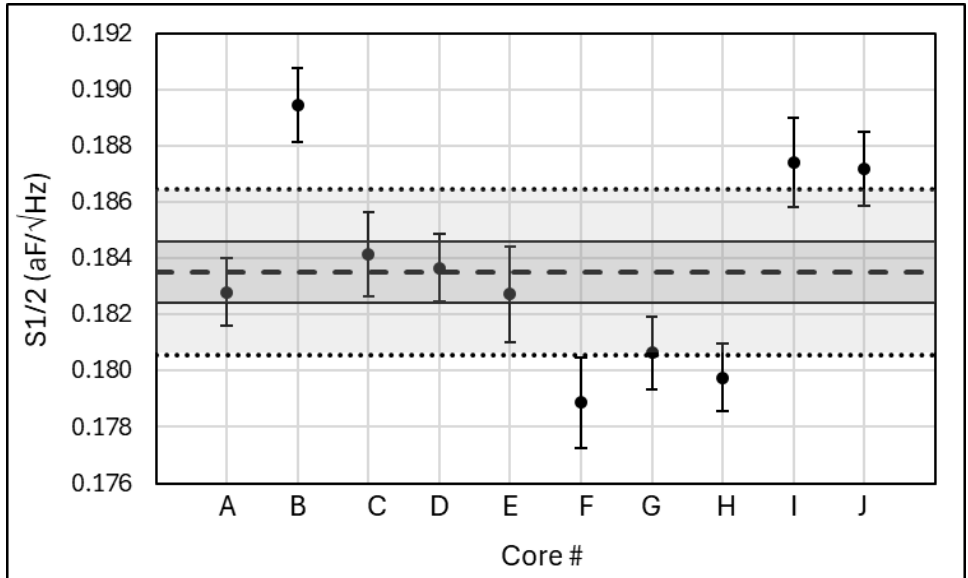

**Figure 10.** 140 K data scaled to 150 kHz with the weighted mean shown by dashed line, variance of weighted mean dark shaded, and 1.5% limits light shaded.

To compare the result from reference 1 to the present work we need to scale both the frequency and the temperature dependence of $S^{1/2}$. Using equation 2:

$$S^{1/2}(f, T) \propto f^{-3/2} L(T)^{-1/2} \propto f^{-3/2} \mu(T)^{-1/2} \tag{9}$$

From reference 11, the permeability $\mu(120\text{K}) \approx 500$ and $\mu(140\text{K}) \approx 600$, thus resulting in: $S^{1/2}(\text{ref }1, 150\text{ kHz}, 140\text{ K}) = 0.172$.

**Table 2 Summary of 140 K data for the A–J cores.**

| Core # | Raw Data | | Parabolic Fit | | | | Scaled to 150kHz |
|---|---|---|---|---|---|---|---|
| | $S^{1/2}_{min}$ (aF/√Hz) | $\Delta S^{1/2}_{min}$ (aF/√Hz) | $f_{min}$ (kHz) | $S^{1/2}_{min}$ (aF/√Hz) | $\Delta S^{1/2}$ (aF/√Hz) | $Q$ | $S^{1/2}_{min}$ (aF/√Hz) |
| A | 0.1794 | 0.0065 | 150.65 | 0.1824 | 0.0012 | 11.2 | 0.1828 |
| B | 0.1810 | 0.0065 | 158.48 | 0.1843 | 0.0013 | 10.9 | 0.1894 |
| C | 0.1811 | 0.0072 | 149.13 | 0.1847 | 0.0015 | 11.0 | 0.1842 |
| D | 0.1769 | 0.0063 | 154.78 | 0.1808 | 0.0012 | 11.2 | 0.1837 |
| E | 0.1861 | 0.0098 | 140.22 | 0.1890 | 0.0017 | 10.9 | 0.1827 |
| F | 0.1765 | 0.0068 | 149.78 | 0.1790 | 0.0014 | 11.3 | 0.1789 |
| G | 0.1734 | 0.0063 | 157.83 | 0.1761 | 0.0013 | 10.9 | 0.1806 |
| H | 0.1687 | 0.0072 | 163.26 | 0.1723 | 0.0012 | 11.0 | 0.1798 |
| I | 0.1842 | 0.0062 | 150.65 | 0.1870 | 0.0016 | 11.0 | 0.1874 |
| J | 0.1829 | 0.0068 | 151.74 | 0.1861 | 0.0013 | 10.9 | 0.1872 |
| **Weighted mean** | | | 152.65 | 0.1815 | | 11.0 | 0.1835 |
| **Variance of weighted mean** | | | 6.34 | 0.0016 | | 0.15 | 0.0011 |
| *Ref. 1 scaled to* | 120 K | | | | | 18 | 0.1276 |
| *150 kHz* | 140 K | | | | | | 0.1720 |

## IV. CONCLUSIONS

In this study we show that, if part of the same manufacturing batch, the noise performance of the RCB sensors is independent of the ferrite cores, at < 0.3% and < 0.6% variance of weighted mean at 293 K and at 140 K respectively. The weighted mean of the RCB minimal noise for the ten cores at 293 K was 0.3027 with a variance of $0.0012\,\text{aF}/\sqrt{\text{Hz}}$ for a frequency range of $85.0 \pm 2.7$ kHz, that, when scaled to 100 kHz resulted in $0.2788\,\text{aF}/\sqrt{\text{Hz}}$ with variance $0.0008\,\text{aF}/\sqrt{\text{Hz}}$. The analogous noise values at 140 K were a weighted mean of $0.1815\text{aF}/\sqrt{\text{Hz}}$ with a variance of $0.0016\text{aF}/\sqrt{\text{Hz}}$ over a frequency range of $152.7 \pm 6.3$ kHz and, when scaled to 150 kHz, a weighted mean of $0.1835\text{aF}/\sqrt{\text{Hz}}$ with a variance of $0.0011\text{aF}/\sqrt{\text{Hz}}$. At 293 K all ten cores are contained in a ± 1% band, while at 140 K nine cores are contained in a ± 1.5% band. Experiments requiring a closer match will of course have to be selected from a larger sample. The frequency dependence of the noise minima is in approximate agreement with the expected scaling of $\propto f_0^{-1/2}$. When scaled for the resonant frequency and temperature these results are consistent with a previous measurement of one other ungapped core, (same type and supplier, different batch) with noise minima agreeing to within $\cong 9\%$ and $\cong 6\%$ at room and low temperature respectively.